\documentclass[]{spie}  %>>> use for US letter paper
\usepackage{amsmath,amsfonts,amssymb}
\usepackage{graphicx}
\usepackage[colorlinks=true, allcolors=blue]{hyperref}

\title{An all-photonic, dynamic device for flattening the spectrum of a laser frequency comb for precise calibration of radial velocity measurements}

\author[a]{Nemanja Jovanovic}
\author[a]{Pradip Gatkine}
\author[b]{Boqiang Shen}
\author[b]{Maodong Gao}
\author[c]{Nick Cvetojevic}
\author[d]{Katarzyna Ławniczuk}
\author[d]{Ronald Broeke}
\author[e]{Charles Beichman}
\author[f]{Stephanie Leifer}
\author[e]{Jeffery Jewell}
\author[e]{Gautam Vasisht}
\author[a,e]{Dimitri Mawet}

\affil[a]{Department of Astronomy, California Institute of Technology, Pasadena, CA, 91125, USA}
\affil[b]{T. J. Watson Laboratory of Applied Physics, California Institute of Technology, Pasadena, CA, 91125, USA}
\affil[c]{Université Côte d'Azur, Observatoire de la Côte d'Azur, CNRS, Laboratoire Lagrange, France}
\affil[d]{Bright Photonics BV, Horsten 1, 5612 AX Eindhoven, The Netherlands}
\affil[e]{Jet Propulsion Laboratory, 4800 Oak Grove Drive, Pasadena, CA 91109, USA}
\affil[f]{The Aerospace Corporation, 2310 E. El Segundo Blvd., El Segundo, CA 90245}

\authorinfo{Further author information: (Send correspondence to Nemanja Jovanovic.)\\Nemanja Jovanovic: E-mail: jovanovic.nem@gmail.com, Telephone: +1 626 395 1214\\ }

\begin{document} 
\maketitle

\begin{abstract}
Laser frequency combs are fast becoming critical to reaching the highest radial velocity precisions. One shortcoming is the highly variable brightness of the comb lines across the spectrum (up to 4-5 orders of magnitude). This can result in some lines saturating while others are at low signal and lost in the noise. Losing lines to either of these effects reduces the precision and hence effectiveness of the comb. In addition, the brightness of the comb lines can vary with time which could drive comb lines with initially reasonable SNR's into the two regimes described above. To mitigate these two effects, laser frequency combs use optical flattener's. 

Flattener's are typically bulk optic setups that disperse the comb light with a grating, and then use a spatial light modulator to control the amplitude across the spectrum before recombining the light into another single mode fiber and sending it to the spectrograph. These setups can be large (small bench top), expensive (several hundred thousand dollars) and have limited stability. To address these issues, we have developed an all-photonic spectrum flattener on a chip. The device is constructed from optical waveguides on a SiN chip. The light from the laser frequency comb's output optical fiber can be directly connected to the chip, where the light is first dispersed using an arrayed waveguide grating. To control the brightness of each channel, the light is passed through a Mach-Zehnder interferometer before being recombined with a second arrayed waveguide grating. Thermo-optic phase modulators are used in each channel before recombination to path length match the channels as needed. 

Here we present the results from our first generation prototype. The device operates from 1400-1800 nm (covering the H band), with 20, 20 nm wide channels. The device was mounted on a PCB board to enable electrical control of the active elements and tested in the laboratory. It was demonstrated that the Mach-Zehnder's allowed for nearly 40 dBs of dynamic modulation of the spectrum, which is greater than that offered by most spatial light modulators. With a smooth spectrum light source (superluminescent light source), we reduced the spectral variation to ~3 dBs, limited by the properties of the components used. On a laser frequency comb which had strong modulations at high spatial frequencies, we still managed to reduce the modulation to ~5 dBs. These devices are of the order of a US quarter and could play a significant role in future PRV and EPRV initiatives.  
\end{abstract}

% Include a list of keywords after the abstract 

\section{INTRODUCTION}
\label{sec:intro}  % \label{} allows reference to this section
\begin{figure}[htbp]
\centering\includegraphics[width = \linewidth]{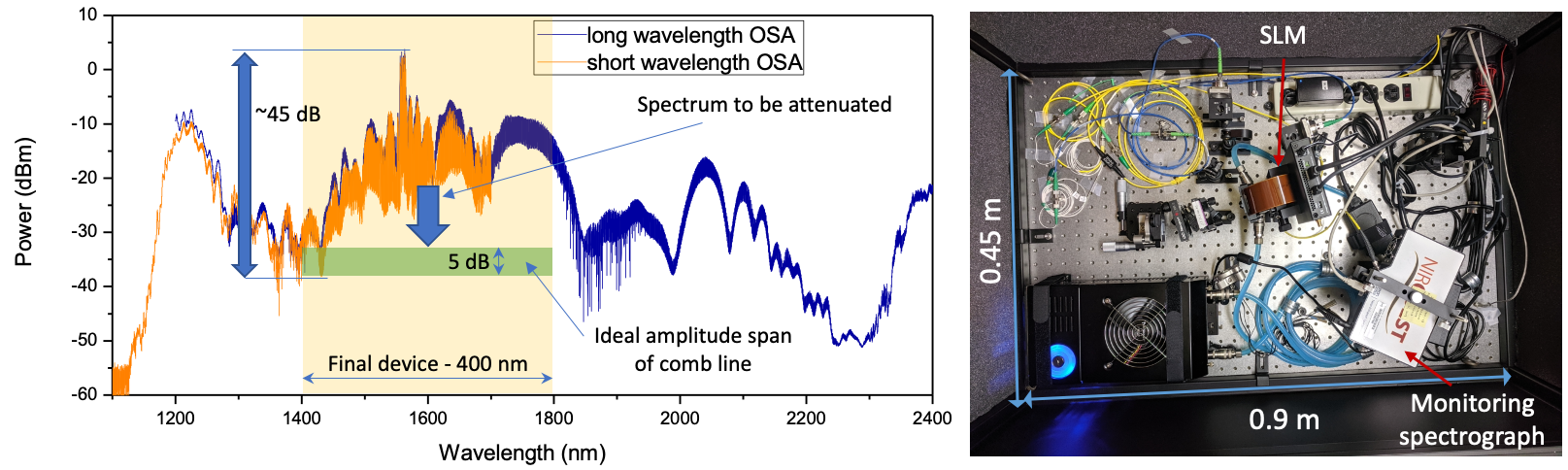}
\caption{(Left) a spectrum of the PARVI LFC showing the strong intensity modulations across the spectrum. (Right) PARVI's bulk optic flattener. The flattener occupies a volume of $1\times0.5\times0.5$ m$^{3}$. \label{fig:bulkflattener}}
\end{figure}

Laser frequency combs (LFC) are becoming essential for the calibration of spectrographs to enable the detection and characterization of exoplanets via the precision radial velocity (PRV) technique~\cite{Fischer2016-EPR}. In order to best utilize an LFC, and hence maximize the accuracy of the wavelength calibration, the largest possible number of comb lines within the observational band pass would be used to derive the wavelength solution. However, the intensity of the lines across the spectrum can vary by many orders magnitude making it impossible to extract all lines with an optimum signal-to-noise ratio (SNR). An example of the issue can be seen in the left panel of Fig.~\ref{fig:bulkflattener}. This shows a trace from the LFC spectrum used with the PARVI spectrograph at Palomar Observatory~\cite{Gibson2019}, which exhibits amplitude differences of up to 45 dB across the 400 nm window, from 1400 to 1800 nm used by the instrument currently. 
To combat this problem, spectral flatteners are typically constructed from bulk optics, where light is dispersed with a grating and a spatial light modulator is used to control the amplitude across the spectrum, before recombining the light into a fiber and sending it to the spectrograph~\cite{probst2013-SFS}. As seen in the right panel of Fig.~\ref{fig:bulkflattener}, these setups can be large (small bench tops) and expensive (several hundred thousand dollars based on two flatteners purchased by this team around the time of this article). Given that the final comb spectrum is typically generated inside a waveguide or optical fiber, finding an all photonic solution to flattening the spectrum would be the most elegant way of miniaturizing this necessary component.

In addition, the brightness of the comb lines can vary with time which could drive comb lines into either the noise floor or non-linear/saturation regions of the detectors dynamic range. The evolution of the amplitude in the spectrum of the LFC is typically slow (on timescales of many minutes to hours) and significantly smaller (1-10 dB) than the amplitude differences across the spectrum. Nonetheless, these variations necessitate that the flattener be a dynamic element that can track and control the evolution in the spectrum with time. 

To address both of these issues, we have developed a broadband, all-photonic spectrum shaper/flattener (BAPSS) on-a-chip. The device was constructed from optical waveguides on a silicon nitride (SiN) chip and operates from 1400-1800 nm (covering the astronomical H band), with 20, 20 nm wide channels. The device was mounted on a PCB board to enable electrical control of the active elements and tested in the laboratory. Here we present the laboratory characterization of a prototype BAPSS device.

%%%%%%%%%%%%%%%%%%%%%%% Circuit design/layout %%%%%%%%%%%%%%%%%%%%%%%%%
\section{Circuit layout}\label{sec:design}
%- outline concept - AWG, MZI, TOPMs, AWG reference previous literature - done
%- Explain how manufacturing has advanced significantly and MPW runs can now be used to great effect. 
%- Outline our requirements are beyond previous work - 400 nm wide - done
%- Show entire PIC layout, explain what devices we made - done
%- Show AWG and sims of spectrum including back-to-back - done
%- Outline MZIs and some MMI optimization and TOPMs - done

\begin{figure}[htbp]
\centering\includegraphics[width = \linewidth]{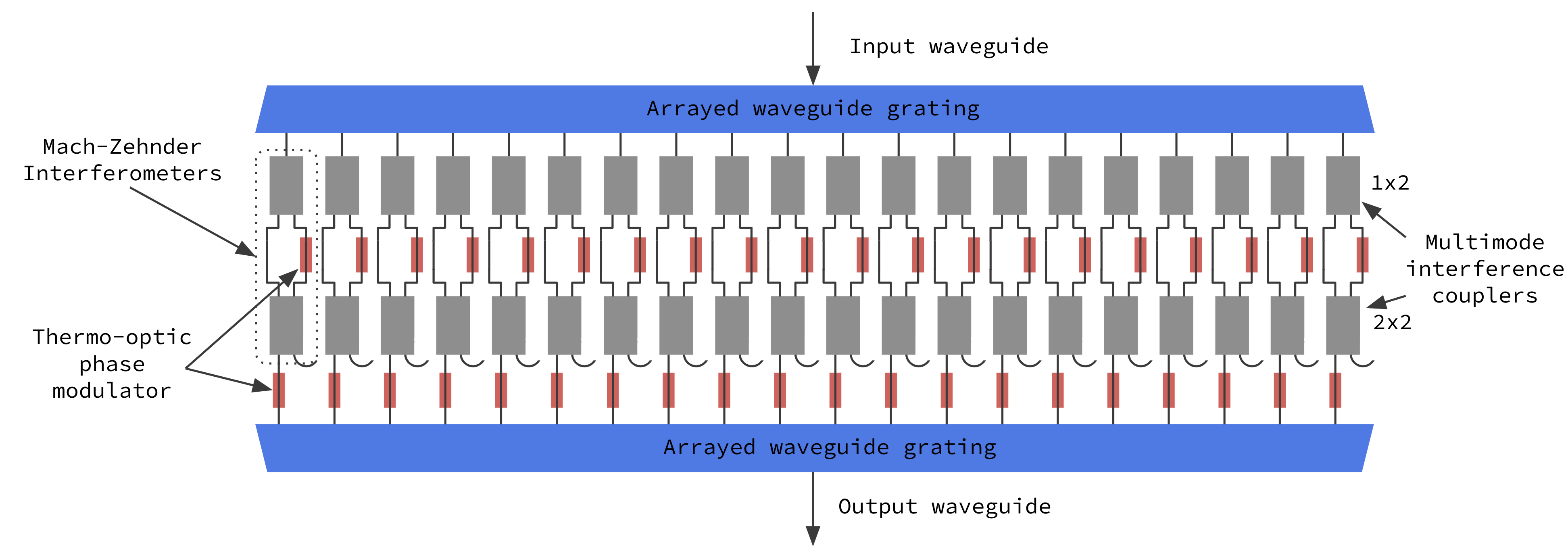}
\caption{Schematic layout of the circuit architecture used for the photonic spectral shaper. \label{fig:concept}}
\end{figure}

The architecture of the circuit we choose for this application is shown in Fig.~\ref{fig:concept} and consists of the following elements. Firstly, an arrayed waveguide grating (AWGs) is used to disperse the light into 20 discrete, narrow band (20 nm) channels. The channel spacing was chosen to match the bandwidth over which the LFC amplitude changes in unison, in addition to reducing the overall number of channels. Next, a series of Mach-Zehnder Interferometers (MZIs), one in each AWG output, are used to adjust the power transmitted through that channel, by modifying the path length on one arm via Thermo-Optic Phase Modulators (TOPMs). TOPMs following each MZI are used to adjust the path length of a given channel with respect to its neighbor to be able to rephase the spectrum. Finally, another AWG with the same properties as the first is used to recombine the channels into a single spectrum. 

This development leveraged significant advances in commercial lithographic fabrication techniques over the last few decades. Indeed, SiN fabrication via Multi-Project Wafer (MPW) runs (fabrication runs where customers pay for only a portion of a wafer), are now producing low loss ($<$0.5 dB/cm at 1550 nm propagation loss) waveguides and devices that are highly reproducible, which was ideal for rapid prototyping of concepts like this. The concept was designed using Nazca Design (\url{https://nazca-design.org/}) with the elements outlined above. 

%In addition to the device, several test structures and circuit elements were also designed to individually characterize the performance of each component. The Photonic Integrated Circuit (PIC) layout is shown in Fig.~\ref{fig:PIClayout}. The image provides an overview with the location and name of various devices and highlights some of the basic elements.

%\begin{figure}[htbp]
%\centering\includegraphics[width = \linewidth]{Figures/Fig3.png}
%\caption{Layout of the PIC with the test devices with key elements labelled. Waveguides are in pin and electrodes and pads in yellow. \label{fig:PIClayout}}
%\end{figure}

%The key elements of the PIC are as follows:
%\begin{itemize}
%    \item A single AWG used to characterize the properties of a standalone AWG (labelled D3).
%    \item A device with back-to-back AWGs (labelled D2) used to measure the spectral reconstruction properties of the AWGs, the relative path length matching of the arms and to see how the spectrum can be modulated in a few channels (1430, 1590, 1770 nm).
%    \item A full BAPSS device which consisted of back-to-back AWGs with power and phase modulators in each MZI arm to test the overall concept (labelled D1).
%\end{itemize}

%%%%%%%%%%%%%%%%%%%%%%% Experiments %%%%%%%%%%%%%%%%%%%%%%%%%

\section{Experimental procedures}\label{sec:experiments}
The circuit was fabricated by LioniX International as part of MPW run $\#$21. The SiN MPW offering consists of a dual stripe waveguide geometry embedded in a silica cladding. For details about the waveguide geometry refer to the MPW overview manual (\url{https://www.lionix-international.com/photonics/mpw-services/}).
%Specifically, the top stripe is 175 nm thick and the bottom is 75 nm thick with a separation of 100 nm. The side walls were inclined at 82$^{\circ}$. A waveguide width of 1.1 $\mu$m was used, 
The waveguides were optimized to allow for optimal guiding around 1550 nm with bend radii as tight as 100 $\mu$m without substantial losses. To improve coupling to optical fibers SSCs were used at the edges of the chip. These consisted of tapers which would expand %ing the top stripe until it vanished. This would expand 
the mode to $\sim$10 $\mu$m around 1550 nm.

\begin{figure}[htbp]
\centering\includegraphics[width = \linewidth]{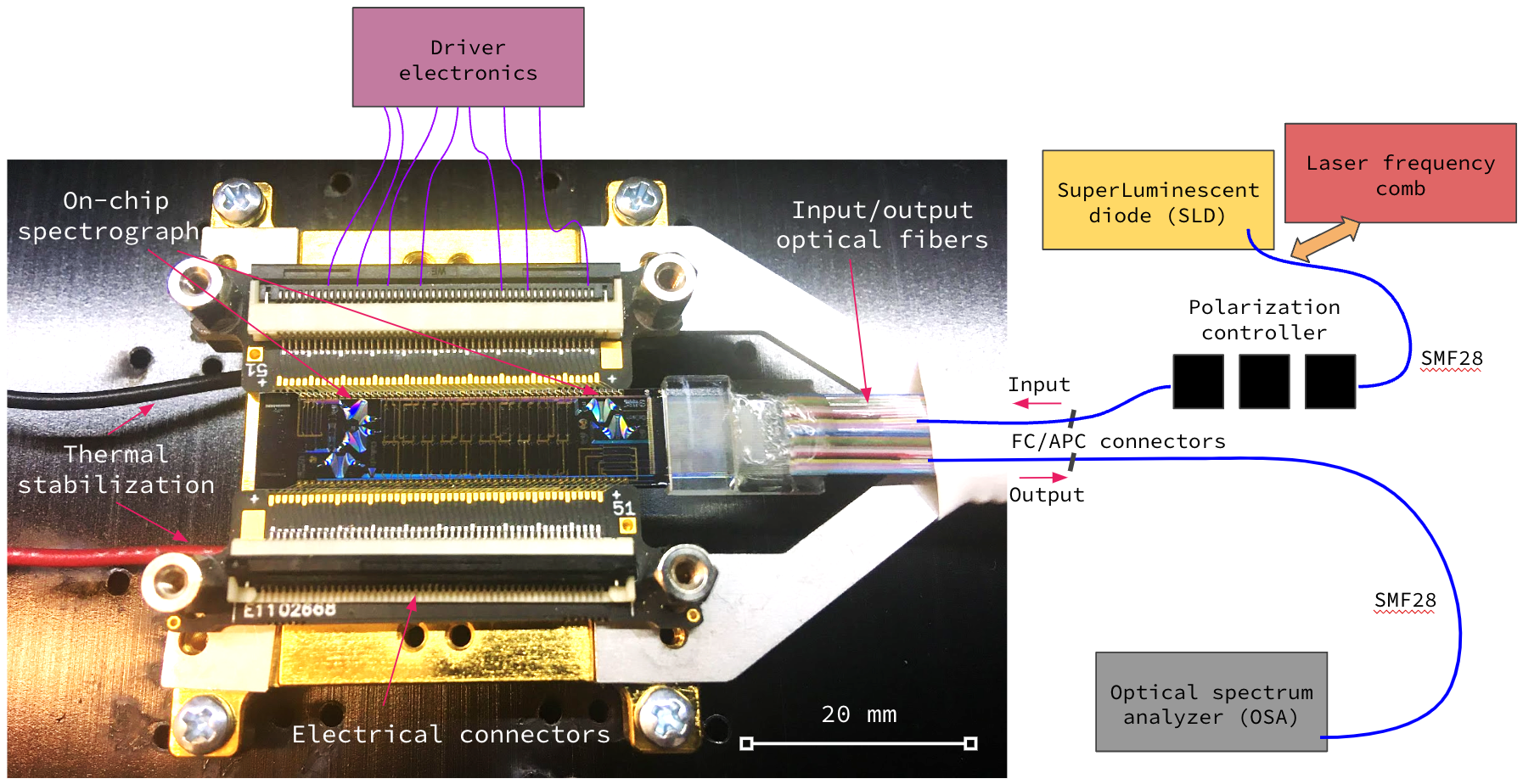}
\caption{Experimental setup for characterizing the broadband all-photonic spectrum shapers. The light source for the tests was either a SLD or an LFC. A polarization controller was used to align the polarization of the source with the TE mode of the waveguides. Light was injected via a v-groove array. The fully packaged PIC is shown. The electrodes on the circuit (chromium plated) as well as the AWGs can easily be seen in the photo. The output was routed to an OSA.   \label{fig:setup}}
\end{figure}

After the devices were fabricated they were diced, polished and packaged to simplify testing. This included bonding a 48-fiber vgroove array to one face of the chip. The fiber used for the vgroove was SMF28. All inputs and outputs to and from the various devices were accessed through this v-groove. The circuit was mounted onto a PCB and the DC and ground lines connected so they could be accessed via ribbon cables from the top and bottom of the device. The entire assembly was mounted on a sub-mount that included a themo-electric cooler (TEC). An image of the packaged device is shown in Fig.~\ref{fig:setup}. 

To test the devices the setup shown in Fig.~\ref{fig:setup} was used. Initial testing was conducted by injecting light from a super-luminescent diode (SLD, Thorlabs, S5FC1550P-A2). Since the fibers bonded to the device under test were SMF28, we used a polarization controller (PCON, Thorlabs, FPC561) to orient the polarized signal of the source with the TE mode of the waveguides on the chip, as the waveguides were highly polarization dependent. The output was connected directly to an optical spectrum analyzer (OSA, Thorlabs, OSA202C).

To test each device, the output of the PCON was connected to a given device, the output of that device to the OSA, and then the PCON was adjusted to maximize the signal on the OSA. This ensured that the polarization of the light source was aligned with the TE mode of the waveguides, which was most efficient. 

To control the TOPMs we used a multi-channel controller. Since the full device consisted of 20 MZIs and 20 TOPMs, the driver had to support $>40$ active channels simultaneously. To achieve a maximum of $3\pi$ phase shifts, we required a driver capable of up to 20 V and 50 mA per line. For this reason we used a multi-channel (120) driver from Nicslab (XPOW-120AX-CCvCV-U). A linear power supply was used to power the multi-channel controller. A computer was used to operate the controller. To connect the relevant pins of the controller to the PCB and hence the device under test, we used an electrical breadboard (Nicslab, M6 multiconnector). 

Once testing with the SLD was completed, we undertook tests with an LFC. However, the comb we had access to was pre-broadened, which meant that it had a triangular shaped spectral profile, not representative of full broadened combs, which are typically flatter, and was only several hundred nanometers wide. Data were acquired with 0.2 nm resolution, where the lines of the comb were not resolved as well as with 0.02 and 0.05 nm resolution where they were.   

The SLD had sufficient flux to test the device down to 1400 nm and up to 1650 nm. The OSA limited testing at the long wavelength end to 1700 nm.

%%%%%%%%%%%%%%%%%%%%%%% Results %%%%%%%%%%%%%%%%%%%%%%%%%
\section{Results}\label{sec:results}
%- Outline results for AWG alone - bandwidth, spacing, etc - done 
%- Show back to back spectrum - done
%- Then full device with different levels of flatteneing
%- Then LFC - flattened at different levels
%- Then talk about power handling and dynamic aspects. 

\begin{figure}[htbp]
\centering\includegraphics[width = 0.8\linewidth]{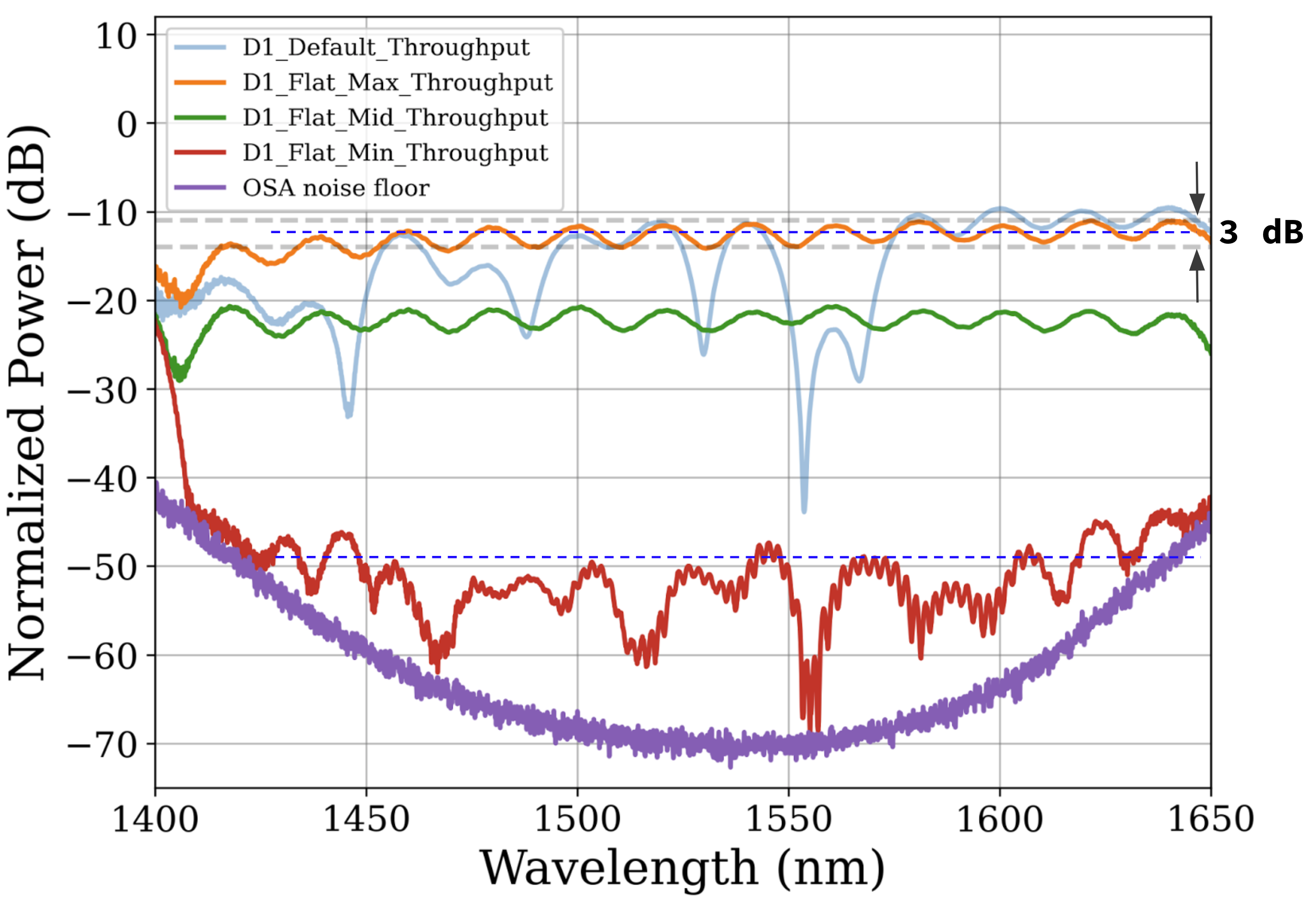}
\caption{Spectral response of the full device. The blue trace shows the transmission of the device with no power applied. The orange is when the MZIs and TOPMS are adjusted for maximum throughput while maintaining flatness. The red trace is when they are adjusted for minimum throughput, and the green for an arbitrary level in the middle of the range. The data were normalized to the reference waveguide/loop back. This process removes the contribution of coupling losses to/from the chip leaving only the losses of the component/device. \label{fig:fulldevice}}
\end{figure}

Using the methods described in the previous section, the spectral response of the BAPSS device was measured. Light was injected into the input of the full BAPSS device and the output routed to the OSA. Spectra were taken with the MZI's turned off (blue) as well as turned on and adjusted for maximum throughput while maintaining flatness (orange), minimum throughput (red) and some arbitrary flat level in between (red) and are shown in Fig.~\ref{fig:fulldevice}. 
The minimum losses occur between 1600-1650 nm, and are as low as 9.5 dBs. The minimum theoretical losses of the AWG and MZI amount to 7.2-7.5 dB, which don't quite account for those measured and could be the result of imperfections in manufacturing over a larger device size (i.e. non-uniformities for example). The spectral ripple on the various flat states (orange and green) have an amplitude of 2-3 dB, which is consistent with what we would expect from the back-to-back AWG. Dashed blue lines indicate the approximate average level for the maximum and minimum throughput states indicating that the power can be adjusted over a 38 dB range. The short and long wavelength ends of the red spectrum were impacted by the noise floor of the OSA (purple trace). 

\begin{figure}[htbp]
\centering\includegraphics[width = \linewidth]{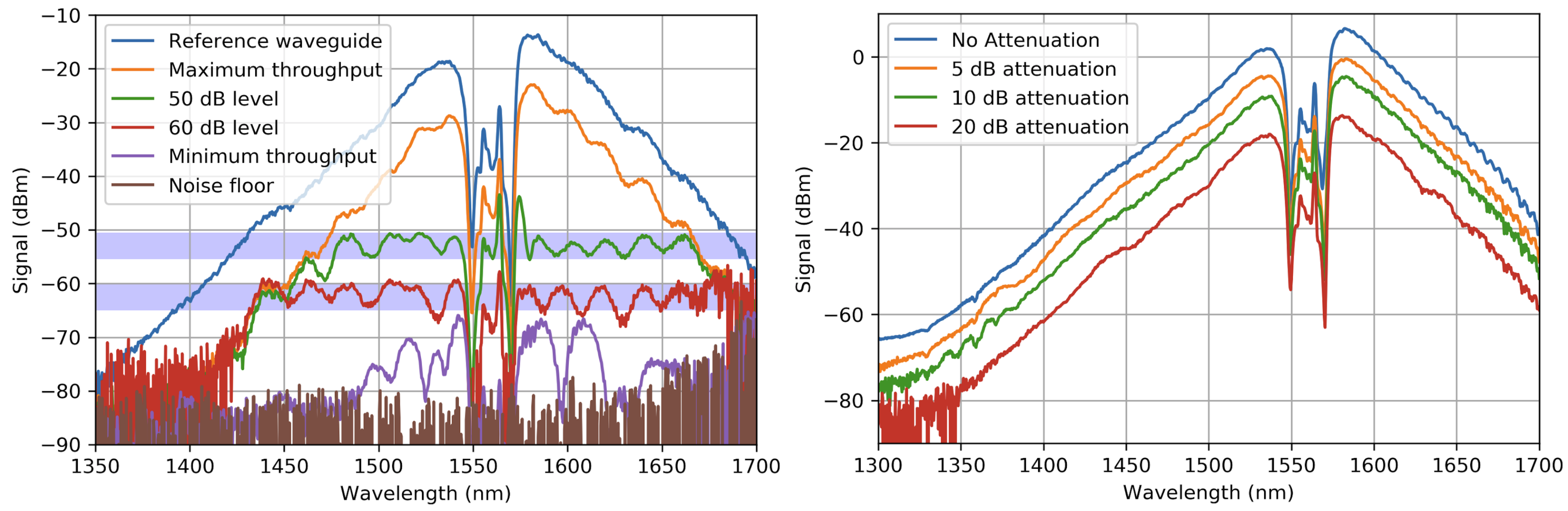}
\caption{Spectral response of the full device with an LFC. (left) LFC spectrum through  the reference waveguide (blue), and through the BAPSS device with maximum throughput (orange), minimum throughput (purple), and two arbitrary flat levels (green and red). The OSA noise floor is shown in brown. (Right) the LFC spectrum taken through the references waveguide with and without optical attenuation. \label{fig:LFCtests}}
\end{figure}

The next set of tests involved the LFC. Figure~\ref{fig:LFCtests} shows the results. The top left panel shows the LFC spectrum through the reference waveguide (blue) as well as the BAPSS device (orange, green, red and purple). It can be seen that the LFC spectrum is triangular in shape on a logarithmic scale, due to the pre-broadened nature of the output of the LFC as outlined in Section~\ref{sec:experiments} above, which is what we had access to. There is a large dip in the region around 1560 nm because a fiber Bragg grating (FBG) was used to suppress the pump region. The grating was not optimized for the spectral profile of the pump or the amplitude and so unfortunately suppressed a lot of light in that region for these tests. 

The orange trace shows the same spectrum through the BAPSS device, with the device set to maximum throughput. The green and red traces show the spectrum flattened at the -50 and -60 dBm levels. The flattening was done manually by eye. The blue shaded boxes are 5 dB high, showing that we managed to flatten the spectrum to within 5 dBs, across hundreds of nanometers of range as was the case with the SLD tests. The only exception to this is the region around the pump, which has a series of sharp spectral features too narrow for the device to control with the current channel spacing's used. Finally, the purple trace shows how much of the LFC spectrum can be suppressed if we put the device into the minimum transmission state. The brown trace is the noise floor of the OSA indicating that in large sections of the spectrum we are limited by the OSA.  

Finally, we tested the power handling capabilities of the device. To ensure we did not damage the device, we connected the LFC to the reference waveguide and took a spectrum. We then replaced the 20 dB attenuator, used in all LFC experiemnts till this point, with a 10 and 5 dB version and finally removed the attenuator altogether. The spectra are shown in the top right panel of Fig.~\ref{fig:LFCtests}. The spectra are largely identical, and are simply offset vertically by the amount corresponding to the attenuation factor. We measured the total integrated power injected into the device with the 20 dB attenuator to be 3-4 mW. As we could not directly measure the power with no attenuator, we can infer it would be a 20 dB brighter signal that is achromatic (all traces are parallel to one another) and therefore assume that in the full power mode, 300-400 mW of power were being injected into the full device. 

After confirming there were no deleterious effects with these levels of power on the reference waveguide, we connected the LFC to the BAPSS device without any attenuation and conducted some basic tests. The BAPSS device performed identically to how it had with reduced power and is therefore capable of working with power levels of typical LFCs. 

Data were also acquired with resolutions of 0.02 and 0.05 nm where the comb lines were resolved, and were deemed to be consistent with the lower resolution data presented here. For simplicity we only show the low resolution data in this paper.

%%%%%%%%%%%%%%%%%%%%%%% Discussion %%%%%%%%%%%%%%%%%%%%%%%%%
\section{Discussion}\label{sec:discussion}
The results above demonstrate that the concept outlined works extremely well for spectral shaping/flattening. Specifically, the device was shown to operate across 250 nm of spectral range, the largest demonstrated from such a layout, with losses consistent with theory. The MZIs were capable of modulating a single spectral channel by 40 dBs, which indicates that the amplitude in the arms of the interferometer is very well balanced, ruling out large manufacturing imperfections. In addition, this modulation range exceeds nearly all bulk optic flatterers which rely on SLMs, which are typically limited to 20-30 dBs only. 

The circuit demonstrated here is actually designed for two purposes: static compensation of the native LFC spectrum as well as for tracking the dynamic aspects. In this work we have mostly focused on compensating for the static profile of the spectrum. It should be noted that around the pump region of an LFC there can be sharp features (as seen in Fig.~\ref{fig:LFCtests}), which would require a photonic device with much narrower channels. This can be done, but more channels means more electrodes and increases the complexity of the circuit. Regardless, the ideal thing to do is to use custom FBGs to compensate for the sharp static aspects of the spectrum, offloading that from the flattener, leaving that to work on the broader parts of the spectrum. If however one wishes to do all of this in the flattener itself, it is possible to use a cascaded AWG approach whereby certain low resolution spectral channels are sent to higher resolution second stage AWGs that provide the sampling needed, but only in a limited number of channels. This is currently being explored and will be focus of future work. 

Although we did not demonstrate the ability of the device to track dynamic changes to the spectrum, the device is more than capable to correct the slow changes expected in an LFC spectrum: with a modulation amplitude of up to 40 dBs and thermal response time as stated by the vendor (and confirmed in a recent publication~\cite{cvetojevic2022-SCS}) of 1 ms it will be sufficient to correct for the slow evolving, low amplitude changes typical of LFC spectra. How to implement the control law for closed loop operation will require some consideration and if the focus of future work.

Astronomical spectroscopy typically requires large bandwidths. We demonstrated successful operation of the BAPSS over a single band (1490-1800 nm) in this work. It may be possible to re-engineer the device to increase the bandwidth to capture several bands and/or split the light at the output of the LFC and send it to several devices each optimized for operation in a different band. This is another avenue that can be explored in future.

\acknowledgments % equivalent to \section*{ACKNOWLEDGMENTS}       
 
This work was produced with support from the Keck Institute for Space Studies. N. Cvetojevic acknowledges the funding from the European Research Council (ERC) under the European Union's Horizon 2020 research and innovation program (grant agreement CoG - 683029).  

% References
\bibliography{report} % bibliography data in report.bib
\bibliographystyle{spiebib} % makes bibtex use spiebib.bst

\end{document}